\newcommand\ion[2]{#1$\;${\scshape{#2}}}%

\documentclass{iau} 
\usepackage{graphicx}

\title[Gaseous abundances in PNe]{Gaseous abundances in planetary nebulae: What have we learned in the past five years?}

\author[Gloria Delgado-Inglada]{Gloria Delgado-Inglada$^1$}

\affiliation{$^1$Instituto de Astronom\'ia, Universidad Nacional Aut\'onoma de M\'exico, 
Apdo. Postal 70264, Ciudad de M\'exico, 04510, Mexico \\ email: {\tt gdelgado@astro.unam.mx}}

\pubyear{2016}
\volume{323} 
\jname{Planetary nebulae: Multi-wavelength probes of stellar and galactic evolution}
\editors{Xiaowei Liu, Letizia Stanghellini, \& Amanda Karakas, eds.}
\begin{document}

\maketitle

\begin{abstract}
Nearly 50 years ago, in the proceedings of the first IAU symposium on planetary nebulae, Lawrence H. Aller and Stanley J. Czyzak said that {\it ``the problem of determination of the chemical compositions of planetary and other gaseous nebulae constitutes one of the most exasperating problems in astrophysics"}. Although the situation has greatly improved over the years, many important problems are still open and new questions have arrived to the field, which still is an active field of study. Here I will review some of the main aspects related to the determination of gaseous abundances in PNe and some relevant results derived in the last five years, since the last IAU symposium on PNe.  

\end{abstract}

\firstsection 
\section{Introduction}
Planetary nebulae (PNe) are the result of the evolution of many stars with masses between $\sim$1 and $\sim$8 M$_\odot$. During their lives, these stars may go through several nucleosynthesis and mixing events that change their initial surface composition and, at the end of their evolution, they eject their outer layers enriching the interstellar medium (ISM) with the newly synthesized products (see, e.g., \cite[Karakas \& Lattanzio 2014]{KL2014}). The PN phase emerges if the previously ejected material becomes ionized by the central star before being diluted in the ISM. The ionized gas of PNe contains the chemical footprints of the former stellar nucleosynthesis and, at the same time, stores information about the chemical composition at the time and place where the progenitor stars were formed. 

PNe constitute important tools to understand the production and evolution of chemical elements in the Universe. The comparison between PN abundances and the predictions from stellar models is very useful to improve our knowledge on the efficiency of nucleosynthesis and transport mechanisms occurring in low- and intermediate-mass stars. The analysis of the element abundances that remain unaltered during the stellar life serves as constraints for galactic chemical evolution models. And the study of the gaseous abundances and depletions of refractory elements in PNe, and their elemental depletions, helps to understand dust formation and evolution in these objects. 

I will review some aspects that should be considered when calculating chemical abundances in PNe and I will highlight some of the main results obtained in the last five years since the previous IAU symposium on PNe. 

\section{Comments on the abundance determination in PNe}

\subsection{The abundance discrepancy factor}
This is one of the most intriguing problems in the field and it remains unsolved. We have known for almost 80 years (\cite[Wyse 1942]{W1942}) that the abundances derived from recombination lines (RLs) are systematically higher than those computed using collisionally excited lines (CELs). The so-called abundance discrepancy factor (ADF), ratio of ionic abundances obtained from RLs to corresponding values from CELs, ranges from 1.5 to 3 in most PNe but reaches $\sim100$ in a few PNe (\cite[Liu et al. 2006]{L2006}; \cite[Corradi et al. 2015a]{C2015a}). 

The abundance discrepancy (AD) may critically affect our knowledge of the evolution of metals in the universe because most of what we know about abundances, especially at large distances, comes from the bright forbidden lines of ionized nebulae. Various scenarios have been proposed to explain this discrepancy, the two preferred ones being the presence of spatial temperature fluctuations (\cite[Peimbert 1967]{P1967}; \cite[Torres-Peimbert et al. 1980]{TP80}) or chemical inhomogeneities in the nebulae (\cite[Liu et al. 2000]{L200}). In both cases, the origin of the inhomogeneities needs to be explained. 

Over the last years several works have been devoted to unravel the reason of this discrepancy and some new ideas have emerged. \cite[Nicholls et al. (2012)]{N2012} suggest that the presence of a significant number of non-thermal electrons in ionized nebulae could affect the determination of temperatures and abundances. The authors are able to explain the differences between the RL and CEL abundances by adopting a kappa distribution for the energies of electrons, instead of the Maxwell-Boltzmann one. However, \cite[Ferland et al. (2016)]{F2016} argue that non-thermal electrons do not survive long enough to affect the collisionally excited lines or the recombination lines. 

Another important contribution to the AD problem has been made by \cite[Corradi et al. (2015a)]{C2015a}. These authors found that the PNe with the most extreme ADFs contain close binary systems. The sample of PNe that satisfies this condition keeps growing (\cite[Jones et al. 2016]{J2016}; \cite[Garc\'ia-Rojas et al. 2016a]{GR2016a}). This finding supports the idea of the existence of chemical inhomogeneities in those PNe, with two different components: one cold ($\sim1000$ K) and metal-rich and another hot ($\sim10000$ K) and with a near solar composition. More discussion on the presence of chemical inhomogeneities in PNe can be found elsewhere in this volume (see articles by M. Pe\~na, J. Garc\'ia-Rojas, and R. Wesson). 

Although the existence of chemical inhomogeneities in some PNe seems doubtless, \cite[Peimbert et al. (2014)]{P2014} claim that this should not be the case for most of the PNe. They find that in 16 PNe for which they can derive reliable electron densities from both CELs and RLs, the densities derived from CELs are similar to or higher than those derived from RLs. They conclude that there is no evidence of oxygen-rich high-density clumps in those PNe, and therefore, that they are chemically homogeneous. 

\subsection{Atomic data}
Atomic data are a key ingredient to derive reliable chemical abundances. Over the past decades, a huge effort has been made by collaborations like the IRON Project\footnote{http://www.usm.uni-muenchen.de/people/ip/iron-project.html} and CHIANTI\footnote{http://www.chiantidatabase.org/} and also by individual scientists, leading to new and more accurate atomic data (a discussion on recent atomic calculations for neutron-capture elements can be found in the article by N. Sterling in this volume). Moreover, softwares, such as PyNeb (\cite[Luridiana et al. 2015]{L2015}), make very easy to explore the effect of different sets of atomic data in the calculations. 

PNe have proven to be excellent laboratories to test atomic data. For example, \cite[Stasi{\'n}ska et al. (2013)]{S2013} compare the values of the densities derived from the observed intensity ratios of [S II] $\lambda$6716/6731 and [O II] $\lambda$3726/3729 for a sample of Galactic PNe with the ratios of the transition probabilities from two sources: \cite[Mendoza \& Zeippen (1982)]{M1982} and \cite[Tayal \& Zatsarinny (2010)]{T2010}. Stasi\'nska et al. find that the oldest atomic data (\cite[Mendoza \& Zeippen 1982]{M1982}) reproduce better the observations and thus, they should be used to compute abundances. This kind of analysis can be easily done to test new atomic data. 

The importance of atomic data in nebular analysis is illustrated by Juan de Dios \& Rodr\'iguez (2017, in prep.; and also in this volume). They find that the differences in the values of O/H and N/O derived using different atomic data can reach up to 0.6 and 1.0 dex, respectively, at densities above $\sim$10000 cm$^{-3}$. They point out that some atomic data should be avoided in the calculations of physical conditions and ionic abundances. 

\subsection{Ionization correction factors}\label{ICFs}
The total chemical abundances can be computed by adding all the ionic abundances of each element. When some ions are not observed, one needs to take into account their contribution through ionization correction factors (ICFs). Originally, the proposed ICFs were based on similarities between the ionization potentials of different ions (e.g., \cite[Peimbert \& Costero 1969]{P69}; \cite[Peimbert \& Torres-Peimbert 1971]{P71}). Later, \cite[Kingsburgh \& Barlow(1994)]{KB1994} compiled a set of ICFs based on around ten photoionization models that were computed to fit the observations of some bright PNe. These ICFs are widely used in the literature but they are found to be inadequate for some elements, such as neon and carbon, since the proposed expressions (Ne/O = Ne$^{++}$/O$^{++}$ and C/O = C$^{++}$/O$^{++}$) are too simplistic (\cite[Delgado-Inglada et al. 2014]{DI2014}). 

\begin{figure}[!h]
\begin{center}
\includegraphics[width=2.7in, trim= 0.5cm 0cm 1.7cm 0cm]{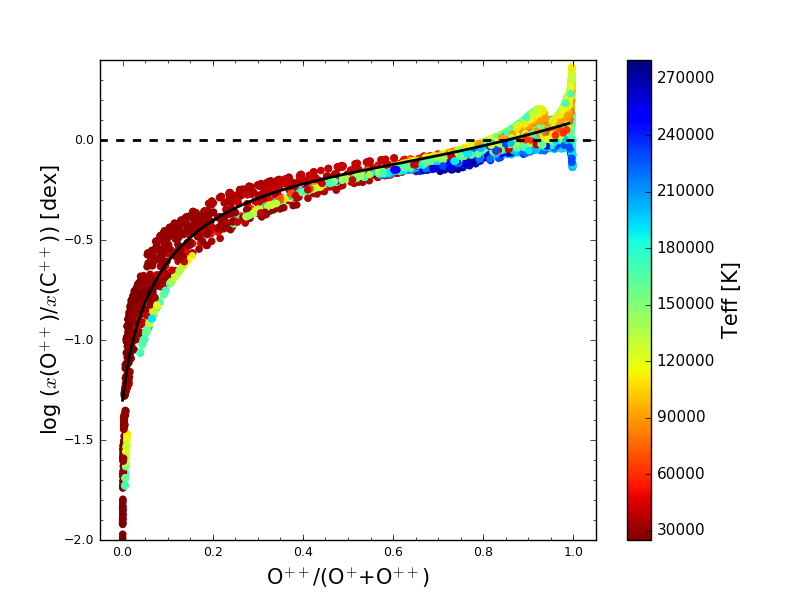} 
\caption{Values of $x$(O$^{++}$)/$x$(C$^{++}$) as a function of O$^{++}$/(O$^{+}$+O$^{++}$) for a sample of photoionization models available in the Million Model Database (\cite[Morisset et al. 2015]{M2015}). The color bar runs from low to high values of the effective temperature. The dashed line represents ICF = 1 while the solid line represents the ICF proposed by \cite[Delgado-Inglada et al. (2014)]{DI2014}.}
\label{fig1}
\end{center}
\end{figure}

This is illustrated in Fig.\ref{fig1} for carbon. The ratio between ionic fractions of O$^{++}$ and C$^{++}$, $x$(O$^{++}$)/$x$(C$^{++}$), is plotted as a function of the degree of ionization for a large sample of photoionization models representative of PNe. The grid is available in the Million Model Database\footnote{https://sites.google.com/site/mexicanmillionmodels/} (\cite[Morisset et al. 2015]{M2015}). Overplotted are the ICF proposed by \cite[Delgado-Inglada et al. (2014)]{DI2014} (solid line) and ICF = 1 (i.e., adopting C/O = C$^{++}$/O$^{++}$; dashed line). It can be seen that whereas for PNe with relatively high degree of ionization, O$^{++}$/(O$^{+}$+O$^{++}$) $>$0.8, both ICFs are similar, at lower degrees of ionization using C/O = C$^{++}$/O$^{++}$ is clearly incorrect (the abundances may be overestimated by as much as 2 dex). 

In the last few years, new ICFs from large grids of photoionization models have become available. \cite[Delgado-Inglada et al. (2014)]{DI2014} provide analytical expressions for the ICFs of He, C, N, O, Ne, S, Cl, and Ar and also for the uncertainties associated to each ICF. Other ICFs recently published in the literature are the ones by \cite[Smith et al. (2014)]{S2014} for Zn, \cite[Sterling et al. (2015)]{S2015} for Se and Kr, and \cite[Delgado-Inglada et al. (2016)]{DI2016} for Ni. 

\section{Results from the study of gaseous abundances in PNe}

\subsection{The sulfur anomaly}
This is one open problem in the field discovered by \cite[Henry et al. (2004)]{H2004} more than ten years ago. It refers to the fact that, at the same metallicity, PNe have systematically lower sulfur abundances than \ion{H}{ii} regions. Nucleosynthesis models do not predict such a destruction of sulfur in the AGB stars (\cite[Shingles \& Karakas 2013]{S2013}). Another option is blaming the ICF for not correcting properly the contribution of unseen ions in PNe (\cite[Henry et al. 2012]{H2012}), but the anomaly is still present when using more modern ICFs to compute sulfur abundances (\cite[Delgado-Inglada et al. 2014]{DI2014}). \cite[Jacob et al. (2012)]{J2012} propose that considering a variable density in PNe (instead of a constant one) may solve the problem. This approach is explored by \cite[Delgado-Inglada et al. 2014]{DI2014} with a grid of photoionization models with a variable density distribution but does not solve the problem. Finally, uncertainties in the dielectronic recombination rates of S$^{++}$ have also been explored without success (\cite[Badnell et al. 2015]{Badnell2015}). In summary, this problem is far from being settled and deserves more study.

\subsection{Oxygen production by low mass stars in the Galaxy}
Oxygen is one of the elements for which more reliable abundances can be calculated in ionized nebulae and its abundance is generally used to trace the metallicity of the interstellar medium at different epochs. This is correct if this element is only produced in massive stars. \cite[Delgado-Inglada et al. (2015)]{DI2015} find evidence of oxygen enrichment (of $\sim$0.3 dex) in PNe of our Galaxy arising from stars of $\sim$1.5 M$_{\odot}$ formed in a subsolar (around half solar) metallicity regime. At low metallicities, it is known that low-mass stars can produce and dredge up oxygen to the surface (\cite[P{\'e}quignot et al. 2000]{P2000}; \cite[Leisy \& Dennefeld 2006]{L2006}) but this is the first observational indication of oxygen enrichment in Galactic disk PNe. 

\begin{figure}[h!]
\begin{center}
\includegraphics[width=2.63in]{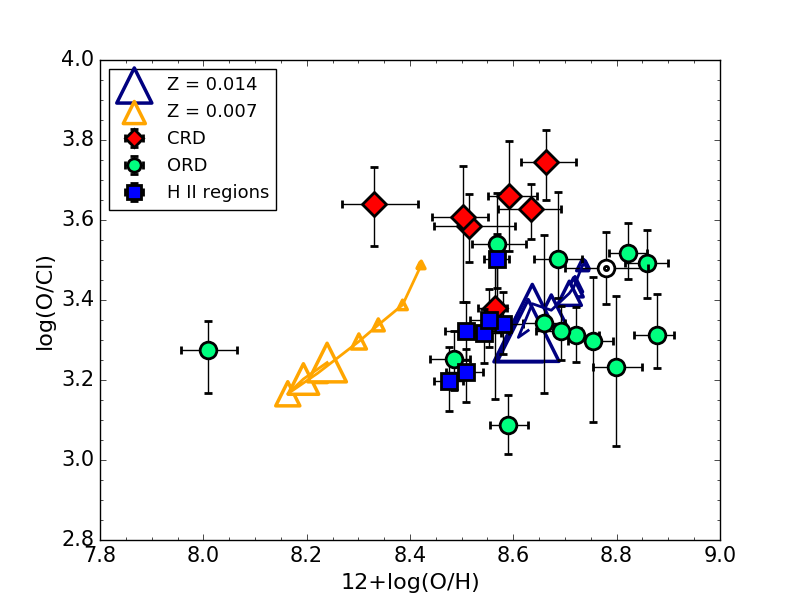} 
\includegraphics[width=2.63in]{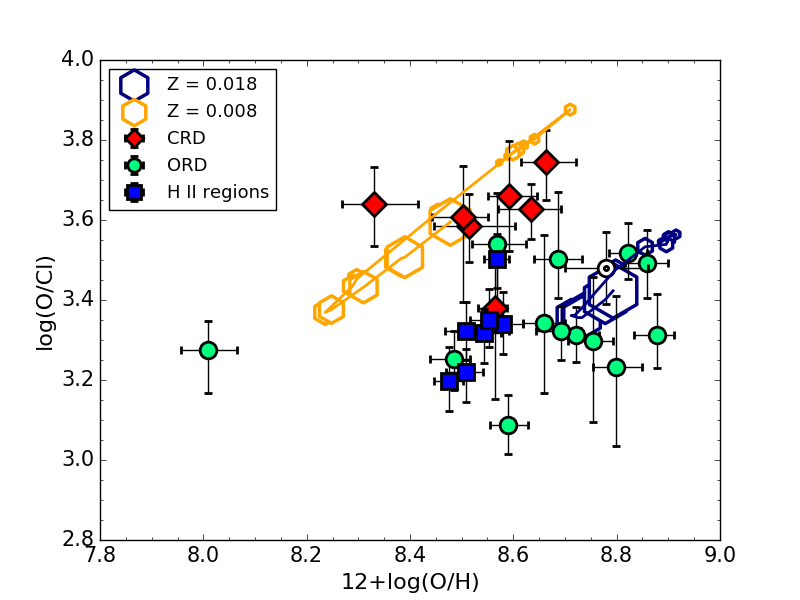}
\caption{Values of O/Cl as a function of O/H for a sample of Galactic PNe with carbon-rich dust (red diamonds) and oxygen-rich dust (green circles) and Galactic \ion{H}{ii} regions (blue squares). Theoretical predictions by \cite[Karakas et al. (2016)]{K2016} (left) and \cite[Ventura et al. (2013)]{V2013} (right) are shown with connected symbols. The solar value from \cite[Lodders (2003)]{L2003} is presented for comparison. The figure has beed adapted from \cite[Delgado-Inglada et al. (2015)]{DI2015}.}
\label{fig2}
\end{center}
\end{figure}

Fig. \ref{fig2} shows the abundances derived for the PNe and \ion{H}{ii} regions studied in \cite[Delgado-Inglada et al. (2015)]{DI2015}. According to the dust features identified in infrared spectra, the PNe are divided into those with carbon-rich dust, CRD, and those with oxygen-rich dust, ORD. The theoretical predictions from \cite[Ventura et al. (2013)]{V2013} and \cite[Karakas et al. (2016)]{V2016} are also plotted in the figures. It can be seen that the models including a more efficient dredge-up, such as the ones by \cite[Ventura et al. (2013)]{V2013}, reproduce better the derived abundances than those with no extra-mixing, such as those computed by \cite[Karakas et al. (2016)]{V2016}.

One important question is how significant is the amount of oxygen produced by low mass stars for the overall enrichment of the Galaxy. The calculations by \cite[Carigi \& Peimbert (2005)]{C2005}, that use the old yields by \cite[Marigo et al. (1996)]{M1996}, \cite[Marigo et al. (1998)]{M1998}, and \cite[Portinari et al. (1998)]{M1998} for low and intermediate stars, predict that only around 1\% of the present oxygen in the solar neighborhood is produced by low-mass stars. New calculations using the yields by \cite[Ventura et al. (2013)]{V2013} predict a somewhat higher production of oxygen ($\sim$2 \%). All the results using different yields will be discussed in a forecoming paper (Delgado-Inglada et al. 2017, in prep.). Although oxygen production by low-mass stars seems to negligible in terms of the overall budget of oxygen in the Galaxy, some caution must be taken when using oxygen as a proxy for the initial composition of PN progenitors.    

\subsection{Neutron-capture elements}
Around half of the elements beyond iron have been produced through slow neutron($n$)-captures in AGB stars (\cite[Lattanzio \& Karakas 2016]{L2016}) but there are still many uncertainties. In particular, it is important to know the efficiency of the mixing mechanism that carries the freshly synthesized elements to the surface and the exact mass at which the source of neutrons changes from the alpha captures of $^{12}$C to the alpha captures of $^{22}$Ne. 

In the last few years, much progress has been done in the study of $n$-capture element abundances in PNe thanks to the combination of high quality spectra, new atomic data, and more reliable ICFs. The abundances of P, F, Ge, Se, Br, Kr, Rb, Ca, and Xe have been determined in several PNe (\cite[Otsuka et al. 2011]{O2011}; \cite[Otsuka \& Tajitsu 2013]{O2013}; \cite[Garc\'ia-Rojas et al. 2015]{GR2015}; \cite[Sterling et al. 2015]{S2015}; \cite[Sterling et al. 2016]{S2016}; \cite[Mashburn et al. 2016]{Mash2016}; Madonna et al. 2017, in prep.). Two of the main important results derived from these studies are: 1) A clear correlation between the C/O values in the PNe and the $n$-capture element enrichment, attributed to the fact that $n$-capture elements and carbon are brought together to the surface; 2) Different amounts of enrichment are found for different $n$-capture elements, and this is thought to be caused by the different progenitor mass. More discussion on the abundances of neutron-capture elements can be found in the articles by N. Sterling and M. Lugaro in this volume. 

\subsection{Ages of stars in the Galactic bulge}
\cite[Nataf (2016)]{N2016} compiled a group of photometric and spectroscopic studies pointing to significantly different ages for the bulge stellar population. Photometric studies found an age of 8 Gyr whereas spectroscopic analysis found a significant population of stars with ages around 3 Gyr. This discrepancy results in the debate on whether the stars in the Galactic bulge are old or not, which has a crucial impact on the idea of how the bulge was formed. It is generally assumed that the bulge is an early product of the Galaxy formation but the presence of a young population would suggest that the formation of the bulge could have happened over a long period of time. 

The analysis of AGB stars and PNe has played a role in this controversy. \cite[Gesicki et al. (2014)]{G2014} studied the star-formation history of the bulge through the analysis of the central star masses (and ages) of 31 bulge PNe. Their results point to an extended star formation in the bulge (i.e. a secular formation of the bulge). The authors concluded that bulge PNe are not representative of the most metal-poor and oldest stellar population. This issue is discussed in more detail by A. Zijlstra in this volume. 

\cite[Buell (2013)]{B2013} find that the observed abundances of bulge PNe, the masses of white dwarfs, and the AGB luminosities, can be well reproduced using models with enhanced N/O and He/H values ($Y \sim0.32$) and initial masses around 1.2--1.8 $M_\odot$ (implying that they were formed between 2 and 5 Gyr ago). The initial enrichment would come from intermediate- and high-mass stars formed earlier in the bulge that contaminated the ISM with their products. The assumption of a pre-enrichment of helium in the bulge would reduce the discrepancy between the photometric and spectroscopic age estimates of the bulge population (\cite[Nataf 2016]{N2016}).  

\subsection{Extragalactic planetary nebulae}
PNe serve to study old stellar populations in external galaxies. Since PNe arise from stars formed at different epochs, their abundances trace the chemical composition of the ISM in the past. If these abundances are compared with those derived from \ion{H}{ii} regions which trace the present day abundances, one can study the chemical enrichment history of galaxies. Besides, the analysis of abundance gradients of PNe and \ion{H}{ii} regions contains information of various physical processes such as gas infall and outflow, star formation history, stellar migration, and the initial mass function. 

A crucial issue before comparing the chemical abundances of extragalactic \ion{H}{ii} regions and PNe is to separate correctly these two types of objects. \cite[Stasi{\'n}ska et al. (2013)]{S2013} provided some useful diagnostics (involving, for example, the H$\beta$ luminosities, N/O values, electron densities and temperatures, and ionized masses) for this purpose.

\cite[Magrini et al. (2016)]{Magrini2016} study the oxygen abundances and gradients of four spiral galaxies, NGC~300, M33, M31, and M81 with PNe and \ion{H}{ii} regions. They find that the global oxygen enrichment increases with time in the four galaxies, being higher for the earlier type ones. They also find that the abundance gradients obtained from the older stellar population (PNe) are similar to or flatter than the ones derived for the younger stellar population (\ion{H}{ii} regions). Other studies of PNe in spiral galaxies are those from \cite[Kwitter et al. (2012)]{K2012}; \cite[Sanders et al. (2012)]{N2012}; \cite[Balick et al. (2013)]{B2013}; \cite[Stasi{\'n}ska et al. (2013)]{S2013}; \cite[Corradi et al. (2015b)]{C2015b}; \cite[Fang et al. (2015)]{F2015}. Some of them are discussed in detail elsewhere in these proceedings. 

Extragalactic PNe can also be used to study stellar nucleosynthesis at different metallicities or in different environments. For example, \cite[Garc{\'{\i}}a-Rojas et al. (2016b)]{GR2016b} used deep and high resolution data obtained with GTC, VLT and CFH to study a sample of 22 PNe in the irregular and low-metallicity galaxy NGC~6822. The authors compare their derived chemical abundances with nucleosynthesis models from different authors to estimate the initial masses of the progenitor stars and find that the predictions differ from one model to another and they also depend on the element that is considered. Moreover, none of the nucleosynthesis models are able to reproduce the helium abundances. This kind of work illustrates that, despite the improvements achieved in observations and models, there is still much work to be done in the field of nucleosynthesis of low- and intermediate-mass stars.  

\subsection{Dust}
The gaseous abundances of refractory elements in PNe provide useful information about the formation and evolution of dust grains in these ionized nebulae, and allow us to explore the different depletion patterns for dust forming elements. 

More than 90\% of the iron atoms present in Galactic PNe are found to be deposited into dust grains (e.g, \cite[Rodr\'iguez \& Rubin 2005]{R2005}; \cite[Delgado-Inglada et al. 2009]{DI2009}). Combining the information of element depletions obtained from the gaseous abundances with the dust composition of the studied objects implied by the infrared dust features, \cite[Delgado-Inglada \& Rodr\'iguez (2014)]{DI2014} find that PNe with carbon-rich dust tend to have higher iron depletions than those of PNe with oxygen-rich dust. This suggests that the formation of dust is probably more efficient in PNe with a carbon-rich environment. Recently, \cite[Delgado-Inglada et al. (2016)]{DI2016} compare the depletion factors of iron and nickel in a group of Galactic PNe and \ion{H}{ii} regions and find that nickel atoms tend to be more attached to the grains than iron atoms at high depletions. This result is interesting because iron and nickel have similar chemical and physical properties and one would expect a similar behavior in their condensation into dust grains. 

\subsection{Final remarks}
I would like to highlight that most of the results discussed here are based on high quality optical spectra, deep and with high resolution. Therefore, it is crucial to obtain more data of such quality, in our Galaxy and other galaxies, to get new and exciting results in the coming years.\\

{\bf Acknowledgements}\\
I thank the organizers of the IAU Symp. 323 for the opportunity to present this work and for the warm hospitaly offered during the conference. I also want to thank the financial support received to attend this meeting from: Peking University, Conacyt grant no. CB-2014/241732, and PAPIIT (DGAPA-UNAM) grant no. 107215.

\begin{discussion}
\end{discussion}

\end{document}